%% file: paper.tex
\documentclass[runningheads]{llncs}
\usepackage[T1]{fontenc}
\usepackage{graphicx}
\usepackage{comment}
\usepackage{amsmath}
\usepackage{tabularx}
\usepackage{array}
\usepackage{multirow}
\usepackage{capt-of}
\usepackage{xcolor}
\usepackage{booktabs}
\usepackage{siunitx}
\usepackage[caption=false,font=footnotesize]{subfig}
\usepackage[disable]{todonotes}
\usepackage{microtype}
\usepackage{xspace}

\usepackage[capitalise]{cleveref}
\usepackage{paralist}
\usepackage{pifont}
\usepackage{scratch3}
\usepackage{url}
\usepackage[most]{tcolorbox}
\usepackage{placeins}
\usepackage{newunicodechar}
\newunicodechar{—}{\textemdash}
\newunicodechar{–}{\textendash}
\newunicodechar{’}{\textquoteright}
\newunicodechar{‘}{\textquoteleft}
\newunicodechar{“}{\textquotedblleft}
\newunicodechar{”}{\textquotedblright}
\newunicodechar{‑}{\nobreakdash-}
\newcommand{\summary}[2]{
	\vspace{0.4em}
	\noindent
	\colorbox{gray!20}{%
		\parbox{.97\linewidth}{%
			\textbf{Summary (\textit{#1})}
			#2
		}%
	}%
}%

\newcommand{\owl}{\textsf{\small CodeOwl}\xspace}

\newcommand{\smallparagraph}[1]{\vspace{0.08em}\noindent {\bf #1}}

\begin{document}
	\title{CodeOwl: Automatic Generation of Tiered Parsons Problems for 
		Introductory Programming}
	\titlerunning{CodeOwl}
	%
	\author{Luca Cisternino\orcidID{0009-0002-6126-3990} \and
		Florian Obermüller\orcidID{0000-0002-6752-6205} \and
		Gordon Fraser\orcidID{0000-0002-4364-6595}}
	\authorrunning{L. Cisternino et al.}
	%
	\institute{University of Passau, Innstraße 41, 94051 Passau, Germany
		\url{https://www.fim.uni-passau.de/en/chair-for-software-engineering-ii}
		 \\
		\email{\{florian.obermueller,gordon.fraser\}@uni-passau.de}}
	\maketitle              
	\begin{abstract}
		Addressing learner heterogeneity in programming education is
		challenging due to variations in student speed, prior knowledge, and
		motivation. While differentiated instruction, such as tiered
		sequences, allows students to engage at appropriate difficulty
		levels, manually creating these resources is labour-intensive.  This
		paper introduces \owl, an AI-driven tool that automates the
		generation of tiered Parsons problems. Starting from a sample task
		or specific programming concepts, \owl produces tiered
		sequences of Parsons problems automatically.
		We evaluated \owl with a mixed-method framework comprising
		complexity analysis, expert ratings, and user studies.  Analysis of
		297 tiered sequences (three tiers each) revealed that $98.7\%$
		achieved a positive complexity increase, successfully rising in
		difficulty from Tier~1 to Tier~3. Experts rated the generated
		problem statements as highly clear. While teachers praised the
		tool's utility, they identified a need for greater control over
		curriculum alignment. Similarly, students reported positively but 
		requested enhanced feedback mechanisms and
		alternative interaction modes.
		
	\end{abstract}
	\input{content/introduction}

	\input{content/background}

\input{content/code_owl}

	\input{content/experimental_setup}

	\input{content/results}

	\input{content/conclusion}

	\begin{credits}
	
	\subsubsection{\ackname} 
	{\small This work was funded
		by INTERREG-Programm Bayern–Österreich 2021–2027 
		BA0100019-Grenzüberschreitender InnRaum³.}
	\subsubsection{\discintname}
	{\small  The authors have no competing interests to declare that are
		relevant to the content of this article.}
	\end{credits}
	%
	%
	%
	\bibliographystyle{splncs04}
	\bibliography{references}
\end{document}

%% file: content/introduction.tex
\section{Introduction}\label{sec:intro}
Learning to program can be challenging and highly depends on the quality of the teaching materials. While teachers may be well versed in the programming concepts they want to teach they also face pedagogical challenges, as they must balance content difficulty, student motivation, and instructional methods to accommodate diverse learners with varying learning speeds, background knowledge, and motivation levels~\cite{bibtex-item-fojcikxchallengesx2022}.
 Computer-science education is particularly 
 heterogeneous~\cite{kloppMarco,waibel2020}, a group of students may already be 
 finished and left without active learning time while another may be struggling 
 to solve the exercise. A highly effective way to tackle this diversity is 
 internal differentiation~\cite{Eikeland02092022,lindner_differentiation_2021}, 
 as it can avoid drawbacks of segregating students into 
 homogeneous groups such as peer stigmatisation~\cite{nonoexternalgrouping}.  An
 example of internal differentiation is a tiered sequence, which offers 
 students tasks of increasing complexity, allowing them to start 
 at their current skill level and progress to more challenging exercises.
 
 However, computer science teachers still struggle with heterogeneous 
 classrooms reporting a persistent lack of differentiation, a core issue in 
 computing education~\cite{bibtex-item-sentancexcomputingx2017}. 
 Differentiation is time-consuming, clashing with the time constraints computer 
 science teachers frequently cite in lesson 
 preparation~\cite{timeConstraintTeacher1,timeConstraintTeacher}. 
 Recent advances in AI, especially Large Language Models (LLMs), have shown 
 potential in automating aspects such as task creation. 
 Prior work has demonstrated that LLMs can generate high-quality 
 multiple-choice questions~\cite{bibtex-item-doughty} and are effective at 
 producing structured tasks such as Parsons 
 problems~\cite{bibtex-item-pratherxrobotsx2023}.
 
 While these approaches focus on generating single exercises, we present 
 \owl, targeting the automated generation of differentiated task 
 sequences. \owl generates differentiated tasks as
 tiered sequences. Each sequence consists of three Parsons problems of increasing difficulty; producing this monotonic difficulty progression across tiers is an explicit design goal of \owl. These are
 automatically generated from a teacher-provided sample exercise within seconds 
 instead of the minutes or even hours it would take to do so manually. We use 
 Parsons problems to reduce cognitive load by eliminating typing and 
 syntax errors, while being as effective as writing 
 code~\cite{ericson2017,smith2023}. Furthermore, to provide instant feedback to 
 students 
 \owl checks correctness of the solution and returns the results.
 To evaluate its effectiveness, we combine quantitative measures with 
 qualitative feedback from teachers and students. 
 In detail, the contributions of this paper are:
 \begin{itemize}
 	\item We propose a framework to automatically create differentiated task 
 	sequences from sample tasks or targeted programming
 	concepts using LLMs.
 	\item We provide a reference implementation of this framework as
 	\owl.
 	\item We report the results of an experiment with 297 generated task 
 	sequences regarding the complexity progression within sequences.
 	\item We show the results of expert ratings of 60 generated problem 
 	statements targetting clarity and redundancy.
 	\item We describe a study with 28 CS educators, collecting ratings of  
 	\owl about the acceptance and usefulness.
 	\item We report on a study with 11 first-semester CS students, collecting 
 	ratings of \owl about task clarity, engagement, workload and 
 	support.
 \end{itemize}
 
 Our results confirm that $98.7\%$ of all generated tiered sequences
 show an increase in complexity going from the first to the third
 tier.  Also, the results of expert ratings suggest that the generated
 problem statements are easy to understand and appropriate for
 beginner level.  Teachers pointed out the usefulness of the tool,
 emphasising the time-saving potential while still generating
 pedagogical valuable tasks.  From the student point of view \owl was
 also seen as supportive and positive for self-efficacy with
 manageable workload.  These are very encouraging findings, however,
 further work is necessary to make the tool more user friendly as
 students wished for stronger feedback while teachers want to have
 more control over the created tasks.

%% file: content/background.tex
\section{Background}\label{sec:background}
Learning to program may be hard, but teachers can support their
students by selecting task formats helping beginners. Parsons problems
have been shown to be particularly helpful for common problems such as
syntax errors and error handling~\cite{atiq2022}. They remove syntax
production while still requiring reasoning about program structure: a
task consists of a student-facing problem statement and a set of
scrambled code blocks that learners must arrange into a correct
order~\cite{bibtex-item-parsonsxparsonsx2006}. This shifts attention
to program structure and the targeted concept, reducing extraneous
cognitive load while still requiring knowledge of control flow and
common coding patterns~\cite{bibtex-item-Sweller1998}. Parsons-style
tasks have been shown to be completed substantially faster without
negative effects on learning outcomes, enabling more practice within
fixed instructional
time~\cite{ericson2017,smith2023,bibtex-item-zhixevaluatingx2019}.
 
Nevertheless, a challenge teachers face when preparing tasks supporting their 
students are highly heterogeneous classrooms: students vary in prior 
knowledge, learning speed, and motivation~\cite{kloppMarco,waibel2020}.
Internal differentiation is a pedagogical approach aimed at meeting the varying 
needs, abilities, and interests of all students, promoting inclusive and 
equitable education, without separating students into different 
classrooms~\cite{bibtex-item-eikelandxdifferentiationx2022,bibtex-item-kathleenxmarkeyxnavigatingx2023}.
 Tiered sequences are one way to implement internal
differentiation~\cite{bibtex-item-10578840,bibtex-item-Richards2007EffectsOT}. 
A tiered sequence is the organisation of a series of tasks in a progression of 
complexity, each building on the previous. Students are free to 
choose at which tier (i.e., difficulty) they start the sequence, allowing for  
self-regulated 
learning~\cite{bibtex-item-article,bibtex-item-tomlinsonxhowx2017}. 
Using Parsons problems can further enhance this effect as they provide 
immediate feedback, maintain student engagement, and effectively scaffold the 
learning process, particularly when tailored to individual learning 
needs~\cite{bibtex-item-Ericson2022Parsons}.
While differentiated tasks are a suitable solution to classroom heterogeneity, 
 lacking availability of such tasks is seen as a central challenge by 
teachers~\cite{bibtex-item-sentancexcomputingx2017,subban_differentiation_2025}.
Time constraints often hinder their practical use because designing well-fitting 
tasks is  
time-intensive~\cite{timeConstraintTeacher1,timeConstraintTeacher}.

Recent advances in large language models (LLMs) offer a promising
avenue to alleviate this burden by automating parts of the task design
process. Consequently, the exploration of ways of introducing LLMs
into education has gained
attention~\cite{bibtex-item-pratherxrobotsx2023}. Research has been
done on LLM’s capabilities to solve code
problems~\cite{bibtex-item-dennyxconversingx2023,bibtex-item-houxmorex2024,bibtex-item-savelkaxcanx2023}
 showing
promising results. Furthermore, LLMs can generate high-quality
multiple-choice questions for programming education that align well
with learning objectives~\cite{bibtex-item-doughty}. These models are
additionally effective at generating structured learning tasks such as
Parsons problems and code
explanations~\cite{bibtex-item-pratherxrobotsx2023}.  However,
exercises created by LLMs often fail to reach an appropriate level
of difficulty~\cite{bibtex-item-delxcarpioxgutierrezxautomatingx2024},
most of the time resulting in exercises far too simple. By creating
tiered sequences, we aim to overcome the challenge of generating
exercises of appropriate difficulty by providing a range of tasks that
increase in complexity. While a tool for generating differentiated
tasks with the help of LLMs has been proposed
before~\cite{bibtex-item-bahrxdifferentiatedx2024}, that tool is
tightly connected to the curriculum of Baden-Württemberg, Germany, and
only creates one task that is differentiated from the input task. Our
approach in \owl creates a full sequence of tiered tasks
in the form of Parsons problems while not being bound to any specific
curriculum.

%% file: content/code_owl.tex
\graphicspath{{content/figures/}}
\section{CodeOwl}\label{sec:codeowl}

\begin{figure}[t]
  \centering
  \subfloat[Instructor input form (filled).%
  \label{fig:homePageFilled}]{
    \includegraphics[width=0.48\linewidth]{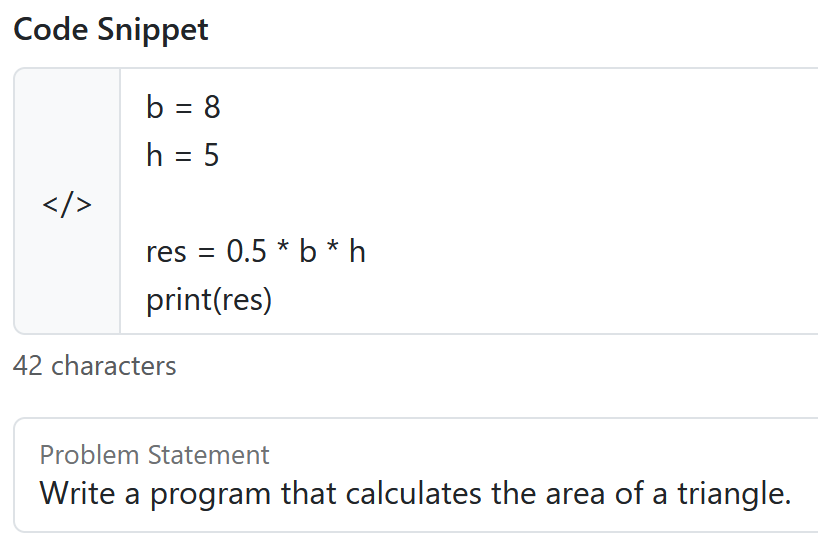}
  }\hfil
  \subfloat[Teacher preview of the tiered sequence.%
  \label{fig:sequenceTeacherView}]{
    \includegraphics[width=0.48\linewidth]{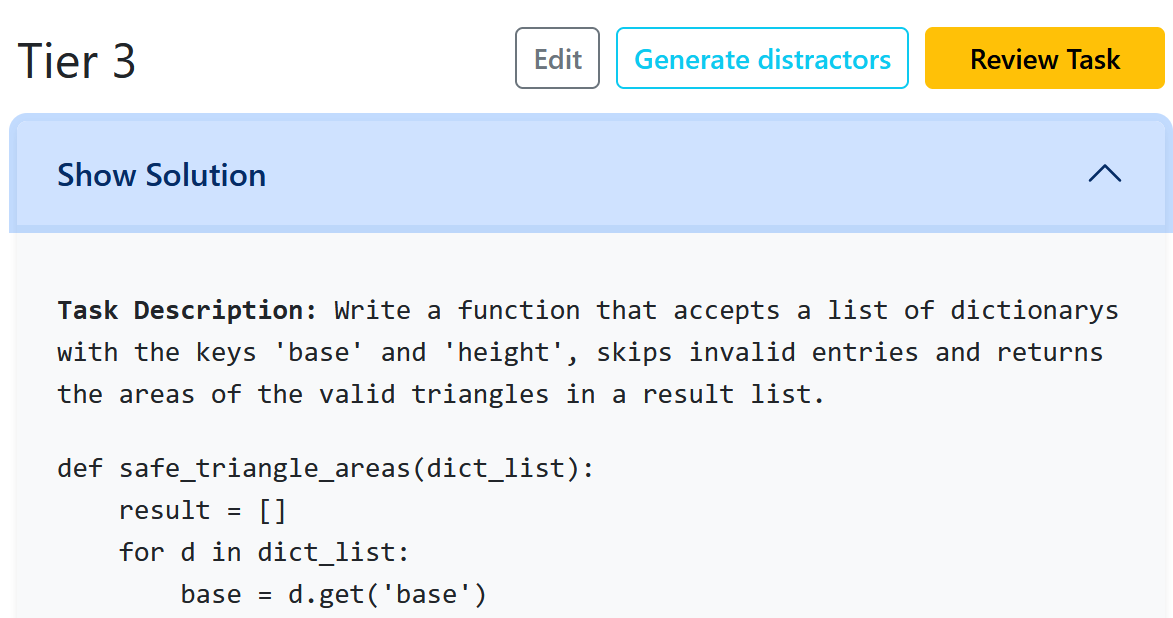}
  }\\[0.6ex]
  \subfloat[Student-facing Parsons task interface (unsolved).%
  \label{fig:taskPage}]{
    \includegraphics[width=0.48\linewidth]{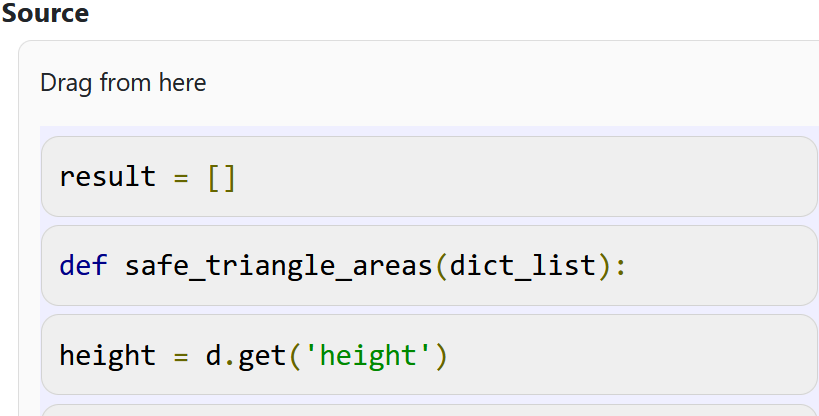}
  }\hfil
  \subfloat[Student-facing Parsons task interface (solved).%
  \label{fig:taskPageSolved}]{
    \includegraphics[width=0.48\linewidth]{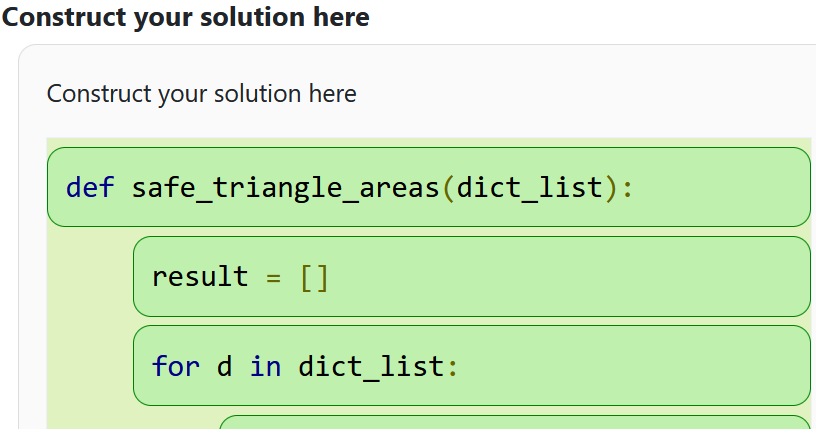}
  }
  \caption{\owl WebUI: (a) teachers provide optional inputs, (b) review tiered tasks and reference solutions in teacher mode, and (c--d) students solve Parsons tasks via drag-and-drop blocks (before and after completion).}
  \label{fig:uiOverview}
\end{figure}

\owl is a web application supporting common classroom use cases such
as differentiated in-class practice (assign tiers by readiness or
allow student choice) and formative assessment (inspect where learners
struggle based on tier completion and errors).

\subsection{Overall Workflow}\label{subsubsec:exampleWorkflow}

Instructors can preview and edit generated content before sharing it
with students, reducing manual authoring overhead while keeping
teacher control.  To generate a tiered sequence they can provide a
\emph{code snippet} (a short, self-contained excerpt of source code),
a problem statement, and optional metadata (\emph{target concepts},
\emph{target language}).
\Cref{fig:homePageFilled} shows an example Python code snippet that
calculates the area of a triangle, along with a fitting problem
statement, Python as the specified language, and two relevant concepts
(cf.~\cref{fig:uiOverview}).
Each field is optional on its own, but either a code snippet or a
problem statement together with a target language must be
supplied. Omitted inputs are automatically generated.

Based on the inputs, \owl outputs a three-tier Parsons sequence with
increasing difficulty, and teachers can preview solutions and curate
content (cf.\ \cref{fig:sequenceTeacherView}).
Difficulty increases across tiers by requiring more complex control
flow, additional parameters, or richer data handling, rather than by
superficial code variation. We fix the number of tiers to three (low,
intermediate, high) as a pragmatic granularity that supports
meaningful in-class differentiation.

Each task consists of a student-facing problem statement and scrambled
code blocks that learners must arrange into the correct order (cf.\
\cref{fig:taskPage}).  \owl tasks provide immediate, block-level
feedback (\cref{fig:taskPageSolved}) during interaction (correct
blocks vs.\ the first incorrect block). Feedback is language-aware:
for indentation-sensitive languages (e.g., Python) indentation
mismatches are treated as errors, while for indentation-insensitive
languages 
indentation is not
graded~\cite{bibtex-item-Ericson2022Parsons}.

\subsection{Tiered Sequence Generation Pipeline}\label{subsec:pipeline}

%


To reliably transform instructor input into a student-ready tiered
sequence, \owl executes a staged generation pipeline with three steps:
%
%
First, the system completes any missing inputs (e.g., inferring a
problem statement or language when needed).

Second, it generates one \emph{contract} per tier---a compact,
structured specification that serves as a uniform interface between
task intent and code generation. Building on empirical evidence that
prompt hygiene and contract-style constraints improve the reliability
of LLM code generation~\cite{siddiq2024faultstarsqualityassessment},
we operationalise this insight via a \emph{solver-facing} contract
(consumed by the code-generating LLM) that fixes the target language,
a one-sentence task statement (describing the problem), an optional
function signature, a brief behaviour specification (optionally via
pre/postconditions), and, when applicable, one to two deterministic
input--output examples. This contract-style separation reduces
ambiguity and increases control over the generated
tasks. 
In practice, tier-to-tier differences in the generated contract are
the main driver of difficulty progression: implementation variations
for the same contract typically have limited impact on task
difficulty, whereas distinct contracts produce different code
structures. Accordingly, we concentrated our prompt optimisation
efforts on generating high-quality tiered contract.
\owl therefore uses prompts with explicit constraints to produce
compact, tiered Parsons tasks. Key design choices include:
\begin{itemize}
    \item \textbf{Explicit monotonicity constraints:} The tier-generation prompt provides complexity levers (e.g., ``+1 decision branch,'' ``add a parameter,'' ``extend from scalar to small collection'') and enforces strict progression.
    \item \textbf{Single-function output:} Solutions are constrained to standalone functions without boilerplate, reducing structural noise and keeping tasks compact. Additionally, the use of code comments is explicitly prohibited.
    \item \textbf{Behaviour-focused problem statements:} A \emph{few-shot} prompt (an LLM prompt containing a handful of worked examples) produces concise, Parsons-compatible instructions that describe observable behaviour without prescribing variable names or implementation details.
\end{itemize}

Third, the system solves each contract, producing one runnable
reference solution per tier. It then converts each solution into a
Parsons task by splitting the code into scrambled blocks and
generating the tier’s student-facing instructions last, conditioned on
the finalised solution to keep the task statement tightly aligned with
the code and optimised for clarity and low redundancy. To reduce
extraneous cognitive load~\cite{bibtex-item-Sweller1998}, statements
are written to admit a single deterministic solution and omit
unnecessary implementation details.


\label{subsubsec:outputPhase}
Before the sequence is exposed to instructors, \owl enforces hard
formatting constraints through a retry mechanism: if a tier’s solution
exceeds a fixed line budget (e.g., 20 lines) or violates output rules
(e.g., contains code comments), the system reruns generation up to
five times and otherwise fails with an explicit error. Once satisfied, the 
sequence is exposed via a
dedicated preview.  The preview allows instructors to inspect each
tier, including reference solutions, and to adapt content if desired.

%

\subsection{System Design and Implementation} 
\owl is implemented as a web application with a Java backend (Spring Boot),
PostgreSQL database, and React TypeScript frontend. The backend isolates
generation logic behind a provider-agnostic AI interface, so we can swap models
and prompt variants without changing pedagogical logic, enabling controlled
experimentation. All sequences analysed in this paper were generated with OpenAI's \texttt{gpt-4.1}.

\subsubsection{Language and Concept Detection}\label{subsubsec:robustness}
Before generating the tiered sequence, \owl detects the target language and programming concepts. When instructors provide a code snippet, concepts are extracted directly from it; when only a problem statement is provided, \owl first generates a minimal reference solution and treats the resulting \emph{concept tags} (labels from the eight-concept taxonomy introduced below) as heuristic metadata.
\begin{itemize}
    \item \textbf{Language Detection:} When the programming language is not explicitly provided, the system detects it from the source snippet using a \emph{zero-shot} prompt (no worked examples). The language detection is conservative: it assigns a language only when there is sufficient evidence, and otherwise falls back to a sensible default rather than guessing. As a result, ambiguous code snippets may not be mapped to a specific language. Instructors should therefore specify the language whenever their snippet could plausibly match multiple languages (e.g., indicate TypeScript if it could also be JavaScript, or Kotlin if it could also be Java). This limitation is not unique to our system—humans also cannot distinguish Kotlin from Java if the snippet could syntactically be Java.

    \item \textbf{Concept Detection:} To align tasks with curricular goals, the system extracts programming concepts from the code using a few-shot prompt. We adopt an intermediate taxonomy that balances the simplicity of Tsai's basic control-flow constructs~\cite{tsai_improving_2019} with the expressiveness of the broader framework described by Luxton-Reilly et al.~\cite{10.1145/3293881.3295779}. Concretely, our taxonomy consists of eight core concepts—\emph{Sequence, Condition, Loop, Variables, Functions, Parameters, Collections,} and \emph{Recursion}—that are educationally meaningful yet feasible for short Parsons problems. This detection step allows the system to tag sequences with metadata, facilitating future retrieval and curriculum alignment.
\end{itemize}

\subsubsection{Task Metadata and Complexity 
Tracking}\label{subsubsec:metadata} 
During generation, \owl computes static code metrics for each tier's reference 
solution—lines of code (LOC), cyclomatic complexity (CC), maximum nesting depth
(MND), method-call count (MCC), and distinct identifier count (DIC)—and
aggregates them into a composite complexity score. We selected these five metrics because they capture complementary, language-agnostic facets of structural complexity in short programs while remaining cheaply computable from a single function. For each metric $k$, we 
z-standardise $x_k$ using language-specific mean $\mu_k$ and standard 
deviation $\sigma_k$:
\[
z_k \;=\; \frac{x_k - \mu_k}{\sigma_k}.
\]
Z-scores are measured in standard deviations from the mean: $z_k=0$ denotes average complexity on metric $k$ within that language, $z_k<0$ indicates below-average complexity, and $z_k>0$ indicates above-average complexity; e.g., $z_k=1$ is exactly one standard deviation above the mean. We standardise per language because languages differ in baseline verbosity and metric distributions, so mixing languages on a single scale would be misleading; z-standardisation also puts metrics on a comparable scale before aggregation. We define composite complexity as the mean of the available standardised metrics
\[
cz \;=\; \frac{1}{K}\sum_{k=1}^{K} z_k,
\]
where $K$ is the number of metrics (in our case $K{=}5$). Higher $cz$ indicates higher code-level complexity relative to other same-language solutions. All metrics are weighted equally, which is a defensible default and often performs on par with expert weightings~\cite{Dawes1979-DAWTRB}. In addition to these metrics, metadata such as target concepts, language, provider, model version, and prompt variant are persisted alongside each sequence to enable comparisons across \emph{generation configurations} (combinations of provider, model version, and prompt variant).


%% file: content/experimental_setup.tex
\section{Experimental Setup}\label{sec:experimental_setup}

We evaluate \owl with a mixed-method design targeting (i) the reliability of tier-wise difficulty progression, (ii)~the classroom readiness of generated problem statements, and (iii)~perceived usefulness and learning experience in realistic use. We triangulate evidence from logged generation outputs (the produced tiered sequences and their stored metadata, cf.\ \cref{subsubsec:metadata}), expert ratings, and teacher and student feedback, and we complement self-reports with click-level interaction traces recorded during student practice: 
\begin{itemize}
  \item \textbf{RQ1:} Do \owl's AI-generated tiered Parsons problem sequences show a clear and meaningful increase in task complexity from T1 to T3?
  \item \textbf{RQ2:} To what extent are the generated problem statements clear and free of redundant information?
  \item \textbf{RQ3:} How do teachers perceive the usefulness of \owl for classroom practice, particularly regarding preparation effort and support for differentiated instruction?
  \item \textbf{RQ4:} How do students perceive and experience \owl's tiered sequences in terms of clarity, engagement, and perceived learning support?
\end{itemize}



\subsection{Datasets and Participants}\label{subsec:setup-samples}

\smallparagraph{RQ1 Dataset: Generated 
Sequences}\label{subsubsec:setup-rq1-data}
A tiered sequence consists of exactly three tiers, \(T1 \rightarrow T2 
\rightarrow T3\), and contains one task per tier (three tasks per sequence). 
The generation dataset consists of tiered sequences (T1--T3) generated 
from \(N_{\mathrm{in}}=99\) base inputs drawn from introductory programming lecture 
material: each base input was anchored by a code snippet taken from the course, 
with the remaining input fields (problem statement, target language, and target concepts)
filled in accordingly. These fields were then selectively included or excluded to 
define three input modes: \emph{full}, \emph{code}, and \emph{meta}.
In \emph{code} mode, the prompt includes a code snippet; in \emph{meta} mode, it includes textual metadata (problem statement, target language, and target concepts); \emph{full} combines both. Each base input is generated in all three modes, yielding a paired (within-input) design for mode comparisons. This results in \(n=99\) sequences per mode (\(3\times 99=297\) sequences total) and \(3\times 297=891\) tasks overall.

\smallparagraph{RQ2 Dataset: Expert Ratings}\label{subsubsec:setup-rq2-data}
Two of the authors, both with a background in computer-science education, assessed the same \(n=60\) tasks, sampled from the \(n=891\) generated
tasks in RQ1 (cf.~\cref{subsubsec:setup-rq1-data}) using a stratified design: 
\(20\) tasks per tier (T1--T3) and \(20\) per input mode (\emph{full}, 
\emph{code}, \emph{meta}). To prevent clustering, sampling without replacement ensured that no two tasks shared a tiered sequence \emph{or} source input.

Raters evaluated each selected task’s problem statement along two dimensions. 
\emph{Clarity} was rated on a 5-point Likert scale with anchored categories from 
\emph{incomprehensible}~(1) 
to \emph{unambiguously clear}~(5), with intermediate levels capturing increasing 
degrees 
of interpretability: clearly ambiguous~(2), ambiguous but directional~(3), and 
clear 
with some reflection required~(4). \emph{Redundancy} was coded as a binary 
indicator capturing whether the statement over-specifies details already fixed by 
the Parsons blocks (e.g., unnecessary low-level instructions).

\smallparagraph{RQ3 Dataset: Teacher Survey}\label{subsubsec:setup-rq3-data}
We surveyed \(N=28\) participants (CS educators and 
teachers in training) after independently creating sequences with \owl. 
Participants were asked to create sequences in any programming language and on any topic, 
and they received only the built-in Help Page as guidance. Based on the 
Technology Acceptance Model (TAM)~\cite{davis_perceived_1989} and extended with 
our domain-specific additions, the instrument measures six constructs on 5-point 
Likert scales: the three core TAM constructs \emph{Perceived Ease of Use} (PEOU), 
\emph{Adoption Intention} (AIN), and \emph{Perceived Usefulness} 
(PU)~\cite{davis_perceived_1989}, plus \emph{Preparation Effort Reduction} (PER), 
\emph{Usefulness for Tiered Differentiation} (UTD), and \emph{Generalisation} 
(GEN). The questionnaire comprises 24 Likert-type items, includes an optional 
free-text field, and captures overall usefulness on a 10-point scale.

\smallparagraph{RQ4 Dataset: Student Study and 
Logs}\label{subsubsec:setup-rq4-data}
We conducted a guided student practice session with \(N=11\) first-semester CS 
majors followed by a questionnaire capturing perceptions of clarity, engagement, 
scaffolding support, self-efficacy, and workload. During solving, \owl logs for every block placement (click) whether it was
correct. Because these logs are not linked to individual students and the number
of clicks varies from task to task, we aggregate them per task rather than per
student and use them only as supporting behavioural evidence (triangulation),
not as a primary outcome.
The practice sessions were offered across two consecutive weeks; 
students could attend either or both sessions at their own discretion. In each week, students worked
through a worksheet in \owl containing sequences generated from that week’s 
introductory 
programming course material. Each session lasted 2 hours and used a worksheet with 10 sequences 
in week 1 and 13 sequences in week 2.

\subsection{Measures and Operationalisations}\label{subsec:setup-measures}

\smallparagraph{RQ1: Complexity Progression}\label{subsubsec:setup-rq1-measures}
For each generated task, we compute a single composite complexity score \(cz\) 
(cf.~\cref{subsubsec:metadata}). To capture tier-wise progression within a 
sequence, we compute pairwise differences in composite complexity between tiers: 
\(\Delta cz_{12} = cz(T2) - cz(T1)\), \(\Delta cz_{23} = cz(T3) - cz(T2)\), and 
\(\Delta cz_{13} = cz(T3) - cz(T1)\). Based on these differences, we derive 
three progression indicators:

\begin{itemize}
  \item \textbf{pass\_basic}: the sequence exhibits an overall increase,
  \(\Delta cz_{13} > 0\).

  \item \textbf{pass\_step}: the sequence shows a substantial increase in at least one step,
  \(\Delta cz_{12} \ge 0.2 \lor \Delta cz_{23} \ge 0.2\).

  \item \textbf{pass\_advanced}: the sequence is strictly increasing across tiers,
  \(\Delta cz_{12} > 0 \land \Delta cz_{23} > 0\), equivalently \(cz(T1) < cz(T2) < cz(T3)\).
\end{itemize}

These indicators are not nested and capture complementary notions of progression strength. To assess the overall tier effect, we report (i) a pooled association between tier index (1--3) and \(cz\) across all tasks (Pearson’s \(r\)) and (ii) a within-sequence view obtained by centring \(cz\) per sequence (subtracting the sequence mean) and recomputing the tier association on the centred values.

Because each of the \(N_{\mathrm{in}}=99\) base inputs is instantiated under all 
three modes (\emph{full}, \emph{code}, \emph{meta}), mode contrasts are analysed 
as paired (within-input) comparisons. For each input \(i\) and progression 
metric,
 we compute paired 
differences between two modes \(A\) and \(B\) (e.g., \(\Delta cz^{(A)}_{12}(i) - 
\Delta cz^{(B)}_{12}(i)\)). 

To quantify the magnitude of paired mode differences, we also report a paired probability-of-superiority effect size \(A_{\mathrm{paired}}\) (stochastic superiority effect size for correlated samples)~\cite{vargha2000}. We report \(A_{\mathrm{paired}}\) because mode contrasts are computed within the same base input, so treating observations as independent would be inappropriate; \(A_{\mathrm{paired}}\) respects the matched design and yields an intuitive probability interpretation. For a comparison labelled ``\(A>B\)'', \(A_{\mathrm{paired}}\) is the probability that a randomly chosen input yields a larger \(\Delta cz\) under mode \(A\) than under mode \(B\), counting ties as \(0.5\):
\[
A_{\mathrm{paired}}=\frac{W + 0.5\,T}{N_{\mathrm{in}}},
\]
where \(W\) is the number of inputs with \(\Delta cz^{(A)}>\Delta cz^{(B)}\), 
\(T\)~the number of ties, and \(N_{\mathrm{in}}=99\). Values near \(0.5\) 
indicate little to no systematic tendency; values \(>0.5\) favour mode \(A\), and values \(<0.5\) favour mode \(B\). This effect 
size is computed on within-input (paired) comparisons; the commonly cited 
Vargha--Delaney \(A_{12}\) refers to the independent-samples 
case~\cite{vargha2000}. For interpretability, we additionally report win/tie/loss 
counts and the median paired difference per contrast. 

This probability-based summary is closely related to the matched-pairs rank-biserial correlation for the Wilcoxon signed-rank test, which likewise contrasts favourable and unfavourable within-pair evidence~\cite{kerby2014}.

To assess whether paired differences are systematically shifted away from zero, we test whether the median paired difference differs 
from zero using two-sided Wilcoxon signed-rank tests; we report statistic \(V\) 
and raw \(p\)-values; zero paired differences are 
treated as ties and omitted from the signed-rank calculation, but retained for 
the win/tie/loss counts and \(A_{\mathrm{paired}}\).
Because we conduct nine tests ($3$ metrics $\times$ $3$ mode pairs), we control 
the family-wise error rate using the Holm--Bonferroni 
adjustment(\(\alpha=0.05\))~\cite{holm1979} .

\smallparagraph{RQ2: Problem Statement 
Quality}\label{subsubsec:setup-rq2-measures}
For \emph{Clarity}, we report both median (IQR) and mean (SD). To quantify uncertainty in the mean clarity score for the small sample (\(n=60\)) without distributional assumptions, we additionally compute a nonparametric bootstrap 95\% confidence interval based on \(B=10{,}000\) resamples (with replacement). Exploratory subgroup summaries by tier and input mode are reported descriptively.
For \emph{Redundancy}, we report the share of tasks labelled redundant. We 
computed interrater reliability on the independent ratings using 
quadratic-weighted Cohen’s \(\kappa_w\) for \emph{Clarity} and unweighted 
Cohen’s \(\kappa\) for \emph{Redundancy}. After reliability estimation, 
disagreements were solved through discussion to obtain one consensus label per 
task.

\smallparagraph{RQ3: Teacher Perceptions}\label{subsubsec:setup-rq3-measures}
For each multi-item construct, we compute a respondent-level construct score as the mean of its corresponding 5-point Likert items. The overall usefulness rating is analysed as a separate single-item measure on a 10-point scale (1--10), independent of the 5-point construct scores (1--5). Analyses are conducted at construct level; respondents with missing values for a construct are excluded from analyses involving that construct. For exploratory relationships between constructs, we compute pairwise Kendall’s \(\tau_b\) correlations on complete cases for the respective construct pair, to identify which perceptions co-vary (in particular with \emph{Adoption Intention}) and to contextualise the construct-level results; correlations are interpreted descriptively rather than causally.

\smallparagraph{RQ4: Student Perceptions and Behavioural 
Signals}\label{subsubsec:setup-rq4-measures}
For each questionnaire construct, we compute a student-level construct score as 
the mean of its corresponding 5-point Likert items. Workload items are 
reverse-coded so that higher scores indicate lower perceived effort. We report 
descriptive statistics per construct and compute pairwise Kendall’s \(\tau_b\) 
correlations on complete cases.

From click-level interaction logs, we compute task-level correctness proxies by 
aggregating all interactions per task. We calculate \emph{share\_correct} as 
the proportion of interactions marked correct and define \emph{wrong-rate} as 
\(1-\textit{share\_correct}\). We use \emph{wrong-rate} for exploratory 
behavioural triangulation by (i) relating it to task complexity \(cz\) via rank 
correlations (Spearman’s \(\rho\), Kendall’s \(\tau\)) and (ii) comparing mean 
task \emph{wrong-rates} across tiers using bootstrap resampling.

\subsection{Threats to Validity}\label{subsec:threats}
Threats to \emph{internal validity} may arise because our evaluation is not a randomised controlled study and evidence partly relies on self-reports, which may be influenced by novelty effects or social desirability. Moreover, the guided nature of the student session and the specific interface may have affected behaviour and perceptions independently of the tiering mechanism; thus, we treat log-based analyses as descriptive triangulation rather than causal evidence.

Threats to \emph{construct validity} may result because we operationalise task difficulty primarily through the composite complexity score \(cz\), which is derived from static properties of generated code and may not capture perceived difficulty or cognitive load during solving. We mitigate single-measure and single-source bias by triangulating across data sources, including complexity progression, expert ratings of statement quality, survey responses, and interaction traces.

Threats to \emph{external validity} may arise because the generation inputs and participants stem from a specific introductory programming context, and results may not generalise across curricula, languages, or populations. In addition, the teacher survey includes heterogeneous stakeholder backgrounds and the student sample is limited in size and setting, constraining generalisability.

Finally, threats to \emph{conclusion validity} may arise from small samples (especially the student study) and multiple exploratory analyses; we therefore emphasise descriptive statistics and effect sizes. Because generation relies on a stochastic LLM (OpenAI \texttt{gpt-4.1}) without a fixed decoding seed, individual outputs are not deterministically reproducible; structured per-tier contracts and a format-checking retry step (cf.\ \cref{subsubsec:outputPhase}) constrain outputs and reject malformed or over-long solutions, so we base conclusions on aggregate patterns across 297 sequences rather than on single generations. Our evaluation also targets perceptions and behavioural proxies rather than learning gains; establishing effects on understanding or skill acquisition would require a controlled pre/post design, which we leave to future work. All respondents consented to data use.


%% file: content/results.tex
\section{Results}\label{sec:results}

\subsection{RQ1: Sequence Complexity Progression}\label{subsec:results-rq1}

\begin{figure}[tb]
  \centering
  \begin{minipage}[c]{0.55\linewidth}
    \centering
    \includegraphics[width=\linewidth]{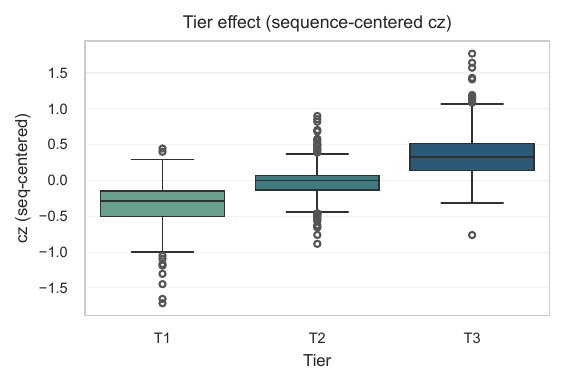}
    \caption{Composite complexity \(cz\) by tier (\(T1\) to \(T3\)) across all
             tiered sequences.}
    \label{fig:rq1-complexity-trend}
  \end{minipage}\hfill
  \begin{minipage}[c]{0.42\linewidth}
    \centering
    \captionof{table}{Pooled shares of sequences meeting progression criteria
                      (\(N_{\mathrm{seq}}=297\)).}
    \label{tab:rq1-pass-global}
    \small\setlength{\tabcolsep}{4pt}
    \begin{tabular}{lc}
      \toprule
      \(N_{\mathrm{seq}}\) & 297 \\
      \midrule
      \emph{pass\_basic} & 98.7\% \\
      \emph{pass\_step} & 86.5\% \\
      \emph{pass\_advanced} & 76.1\% \\
      \bottomrule
    \end{tabular}
  \end{minipage}
\end{figure}

Across tiered sequences, composite complexity increases from \(T1\)
to \(T3\), indicating that \owl can induce a tier-wise difficulty
progression at scale. This upward trend is visible both in pooled
analyses and when controlling for per-sequence baselines.
\Cref{fig:rq1-complexity-trend} shows a clear upward shift in \(cz\) from 
\(T1\) to \(T3\). In a pooled view across all tasks, \(cz\) correlates 
positively with tier index (Pearson’s \(r \approx 0.42\), \(p<.001\)). When 
centring \(cz\) within each sequence—thus isolating within-sequence changes and 
removing between-sequence baseline differences—the association becomes 
substantially stronger (Pearson’s \(r \approx 0.77\), \(p<.001\)).

Using the pass indicators defined in \cref{subsubsec:setup-rq1-measures}, progression holds for a large majority of sequences. \Cref{tab:rq1-pass-global} summarises pooled shares across all analysed sequences (\(N_{\mathrm{seq}}=297\)). Nearly all sequences satisfy \emph{pass\_basic} (98.7\%), indicating an overall increase from \(T1\) to \(T3\). A large share also meets \emph{pass\_step} (86.5\%), meaning that at least one tier transition shows a substantial jump in complexity. Finally, 76.1\% of sequences are strictly increasing across all tiers (\emph{pass\_advanced}). Among the remaining sequences that are not strictly increasing, most still increase overall but exhibit a local plateau or reversal in one step.

Across the nine paired mode contrasts (three \(\Delta cz\) metrics \(\times\) 
three mode pairs), we find no statistically detectable mode effect on 
progression deltas: all Wilcoxon signed-rank tests are non-significant after 
Holm correction (all \(p_{\mathrm{holm}}=1\); \cref{tab:rq1-mode-comparisons}). 
Consistent with this, effect sizes are small: \(A_{\mathrm{paired}}\) values 
cluster near \(0.5\) and median paired differences are close to zero across 
contrasts. The largest deviation occurs for \(\Delta cz_{12}\) (meta vs.\ code; 
\(A_{\mathrm{paired}}=0.404\), median difference \(=-0.034\)), suggesting a 
slight tendency for \emph{code} to yield larger first-step gains, but this 
pattern is not statistically significant. Overall, mode choice does not
measurably affect the strength of tier-wise complexity progression.
\begin{table}[tb]
\centering
\small
\caption{Paired mode comparisons on per-input progression deltas (\(N_{\mathrm{in}}=99\)). \(A_{\mathrm{paired}}\) is the paired probability-of-superiority; \(\widetilde{\Delta}\) is the median paired difference (first mode minus second). Wilcoxon signed-rank tests are two-sided; \(V\) denotes the signed-rank statistic; \(p_{\mathrm{holm}}\) applies Holm correction across all nine tests.}
\label{tab:rq1-mode-comparisons}
\setlength{\tabcolsep}{4pt}
\begin{tabular}{ll S[table-format=1.3] S[table-format=-1.3] S[table-format=4.1] S[table-format=1.3] S[table-format=1.0]}
\toprule
\emph{Metric} & \emph{Comparison} & {$A_{\mathrm{paired}}$} & {$\widetilde{\Delta}$} & {$V$} & {$p$} & {$p_{\mathrm{holm}}$} \\
\midrule
\(\Delta cz_{13}\) & meta $>$ full & 0.480 & 0.000 & 2139.0 & 0.724 & 1 \\
                  & full $>$ code & 0.485 & -0.011 & 2312.0 & 0.816 & 1 \\
                  & meta $>$ code & 0.530 & 0.005 & 1966.5 & 0.316 & 1 \\
\midrule
\(\Delta cz_{12}\) & meta $>$ full & 0.434 & -0.003 & 1702.5 & 0.371 & 1 \\
                  & full $>$ code & 0.455 & -0.007 & 2078.0 & 0.953 & 1 \\
                  & meta $>$ code & 0.404 & -0.034 & 1832.0 & 0.302 & 1 \\
\midrule
\(\Delta cz_{23}\) & meta $>$ full & 0.515 & 0.007 & 2132.0 & 0.379 & 1 \\
                  & full $>$ code & 0.500 & 0.000 & 2196.5 & 0.631 & 1 \\
                  & meta $>$ code & 0.510 & 0.003 & 2137.5 & 0.307 & 1 \\
\bottomrule
\end{tabular}
\end{table}

\summary{RQ1}{Composite complexity generally increases from \(T1\) to \(T3\), indicating that \owl can induce tier-wise difficulty progression at scale. This progression holds for the vast majority of sequences, and input mode does not meaningfully change the strength of the progression.}

\subsection{RQ2: Problem Statement Quality}\label{subsec:results-rq2}

\begin{figure}[!tb]
	\centering
	\includegraphics[width=0.8\linewidth]{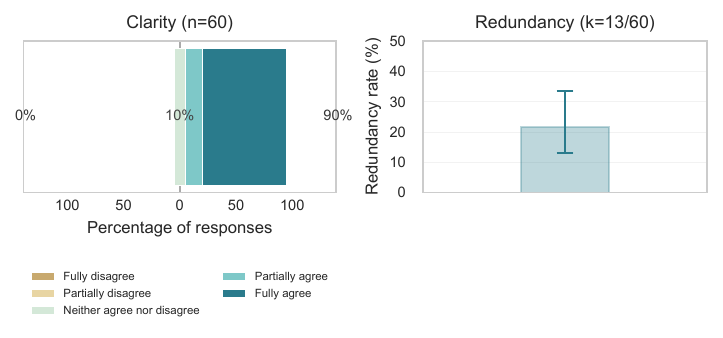}
	\caption{RQ2—Expert ratings for problem statements (\(n=60\) tasks). Left: 
	clarity ratings on a 5-point scale anchored at \emph{incomprehensible} (1) 
	and \emph{unambiguously clear} (5). Right: redundancy rate; 
	\(k=13/60\) tasks were rated redundant.}
	\label{fig:rq2-expert-ratings}
\end{figure}

\begin{figure}[!tb]
	\centering
	\includegraphics[width=0.95\linewidth]{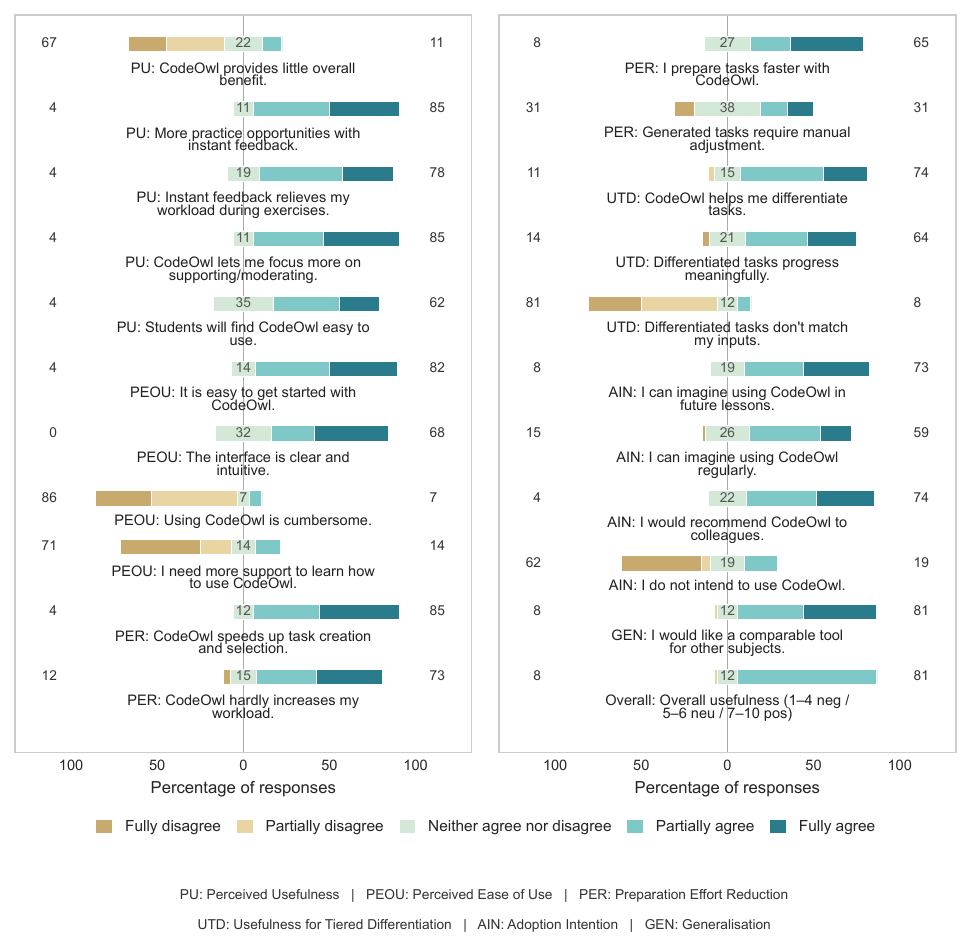}
	\caption{Response distributions for survey items from teachers (\(N=28\)). 
		Disagreement is shown left, agreement right, and neutral 
		responses are centred at zero.}
	\label{fig:rq3}
\end{figure}

\begin{figure}[tb]
	\centering
	\includegraphics[width=\linewidth]{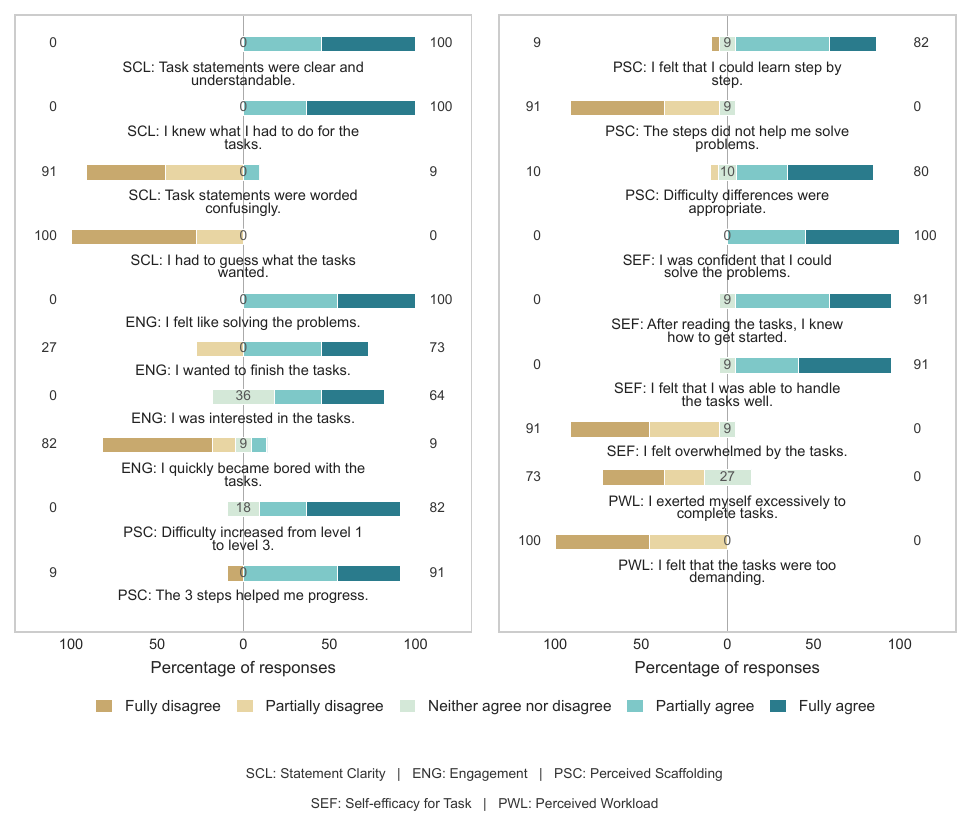}
	\caption{Response distributions for survey items from students (\(N=11\)). 
		Disagreement is shown left, agreement right, and neutral 
		responses are centred at zero.}
	\label{fig:rq4}
\end{figure}

Overall, expert ratings indicate that the generated problem statements are clear and appropriate for the intended level, with redundancy present only in a subset of tasks.
As shown by \cref{fig:rq2-expert-ratings}, clarity was rated very high across tasks (\(n=60\)), with a mean of \(\bar{x}=4.65\) (\(SD=0.66\)) and a median of \(5.0\) \([Q_1=4.75,\,Q_3=5.0]\). The bootstrap 95\% confidence interval for the mean was \([4.48,\,4.82]\). Descriptively, clarity remained high across tiers and input modes.

Overall, 21.7\% of tasks were flagged as redundant (cf.\ \cref{fig:rq2-expert-ratings}). Redundancy typically reflected minor over-specification (e.g., explicitly naming variables or methods) and was rarely excessive. We observed no systematic relationship between redundancy and clarity.
Inter-rater agreement was \(\kappa_w=0.66\) for clarity (quadratic-weighted) and \(\kappa=0.75\) for redundancy, corresponding to \emph{substantial} agreement under common interpretation guidelines~\cite{landis1977measurement}.

\summary{RQ2}{Expert ratings indicate that generated problem statements are very clear and appropriate for the intended level. Redundancy appears only occasionally and is typically mild, and rater agreement suggests the quality assessment is robust.}

\subsection{RQ3: Teacher Perceptions}\label{subsec:results-rq3}
Overall, responses were consistently positive: all constructs were above the 
neutral midpoint (3 on the 5-point scale), and the separate overall usefulness 
item (10-point scale) was high, as shown in \cref{fig:rq3}.
 \emph{PEOU} and \emph{PU} were highest (\(M=4.10\) and \(M=4.02\)), with 
 \emph{PU} showing the strongest consensus (\(SD=0.53\)). The domain-specific 
 constructs were also clearly positive (\emph{UTD}: \(M=3.89\), \emph{AIN}: 
 \(M=3.88\); both medians \(=4.00\)), while \emph{PER} showed more dispersion 
 (\(SD=0.84\), \(IQR=1.38\)), suggesting heterogeneous time-saving experiences. 
 \emph{GEN} had the highest mean among the 5-point items (\(M=4.15\)), 
 indicating strong interest in transferring the differentiation approach beyond 
 programming. Overall usefulness on the 10-point scale was high (\(M=7.77\), 
 \(median=8.00\)). No systematic association emerged between prior AI/EdTech 
 experience and overall usefulness.

Pairwise Kendall’s $\tau_b$ correlations suggest that willingness to adopt is most closely linked to perceived \emph{pedagogical and practical value}: \emph{Adoption Intention} aligned most strongly with \emph{UTD} ($\tau_b \approx 0.60$), \emph{PU} ($\tau_b \approx 0.55$), and \emph{PER} ($\tau_b \approx 0.54$). In contrast, \emph{PEOU} was rated very highly but showed weaker associations with other constructs (e.g., $\tau_b \approx 0.17$ with PER; $\tau_b \approx 0.08$ with UTD), suggesting that ease of use was more of a baseline expectation than a key differentiator of overall judgments. 
Open-ended comments ($n=21$) corroborated the survey pattern: respondents frequently described \owl as fitting typical lesson preparation workflows and enabling quick generation of tiered practice, while noting concrete improvement needs. The most common themes concerned \emph{UI/interaction details} (especially drag-and-drop friction for longer tasks and clearer completion feedback), and \emph{teacher control and documentation} (examples of good inputs, restricting constructs, limiting code length, and better curriculum alignment). Instructors perceived the core idea as valuable and transferable (GEN), while highlighting practical steps required for classroom-ready use.

\summary{RQ3}{Teachers rated all constructs positively and considered \owl useful. Willingness to adopt aligns most with perceived pedagogical/practical value and time-saving potential, while ease of use is a baseline expectation rather than a differentiator; comments emphasise smoother UI interaction and more teacher control and guidance.}

\subsection{RQ4: Student Perceptions and Behavioural Triangulation}\label{subsec:results-rq4}

%
%

\begin{figure}[tb]
  \centering
  \subfloat[Wrong-rate vs.\ composite complexity \(cz\) (points coloured by tier; size \(\propto n_{\text{clicks}}\)). LOWESS trend shown; higher \(cz\) is associated with higher wrong-rate (Spearman \(\rho=0.62,\, p=0.0013\); Kendall \(\tau=0.46,\, p=0.0020\); \(N=24\)).%
  \label{fig:rq4-wrongrate-vscz}]{
    \includegraphics[width=0.50\textwidth]{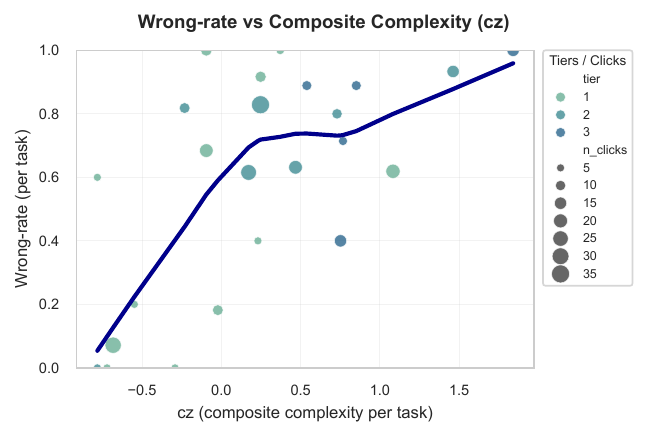}
  }\hspace{0.02\textwidth}
  \subfloat[Bootstrap distributions of mean task wrong-rate by tier (\(B=10{,}000\)); dots mark observed means. Wrong-rate increases from Tier~1 to Tier~3.%
  \label{fig:rq4-wrongrate-bytier}]{
    \includegraphics[width=0.37\textwidth]{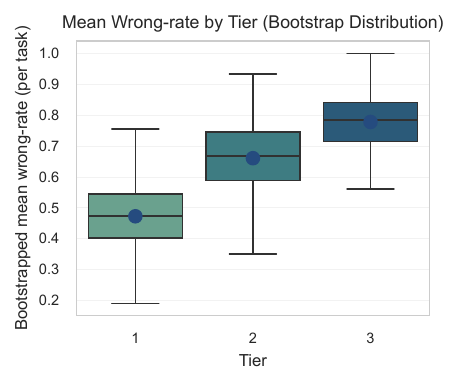}
  }
  \caption{Task error patterns in the feedback logs.}
  \label{fig:rq4-feedback-wrongrate}
\end{figure}

Students (\(N=11\)) rated all constructs above the neutral midpoint. \emph{Statement Clarity} was highest (\(M=4.55\), \(SD=0.63\)) and \emph{Self-efficacy} was also high (\(M=4.41\), \(SD=0.62\)). \emph{Engagement} and \emph{Perceived Scaffolding} were positive but more heterogeneous (\(SD=0.96\) each). Workload (higher \(=\) lower effort) suggests manageable effort rather than overload (\(M=4.32\), \(SD=0.72\)) (cf.~\cref{fig:rq4}). 
Exploratory correlations (Kendall’s $\tau_b$, computed on complete cases) suggest a coherent self-report pattern: clarity aligns with self-efficacy ($\tau_b\approx 0.59$) and self-efficacy aligns with lower workload ($\tau_b\approx 0.57$), consistent with the interpretation that clear statements support confidence and reduce perceived effort.
Mean wrong-rates increase from Tier~1 to Tier~3 in the aggregate (cf. \cref{fig:rq4-wrongrate-bytier}), consistent with a difficulty progression. However, within-sequence trajectories are heterogeneous: many sequences increase cleanly across tiers, while others are flatter or show local reversals (e.g., Tier~3 below Tier~2).
\Cref{fig:rq4-wrongrate-vscz} relates task difficulty to observed performance: tasks with higher composite complexity \(cz\) tend to exhibit higher wrong-rates in the feedback logs. This association is substantial and consistent with \(cz\) capturing task difficulty (Spearman \(\rho=0.62,\, p=0.0013\); Kendall \(\tau=0.46,\, p=0.0020\); \(N=24\) tasks).
At the same time, the task-level signal is noisy and varies across sequences; given (\(N=11\)) and non-student-indexed logs, we treat this triangulation as exploratory
open-ended responses ($n=11$) corroborated the survey pattern. Students most 
often framed \owl as \emph{useful for getting more practice} and as a good fit 
for regular course routines (e.g., short warm-ups), while emphasising that 
reuse depends on trusting snippet correctness. The most common usability 
request concerned trackpad friction: students found drag-and-drop effortful and 
asked for click-based placement. A smaller but notable theme concerned 
\emph{strict evaluation}: students reported rejection of valid alternative 
orderings, as the interface checks only one reference solution.

\summary{RQ4}{Students reported high clarity, supportive tiering, and strong self-efficacy with manageable workload. Click-level logs point to increasing difficulty across tiers and higher error rates for structurally more complex tasks, while open-ended feedback highlights drag-and-drop friction and overly strict checking of alternative correct solutions.}


%% file: content/conclusion.tex
\section{Conclusions}\label{sec:conclusions}

In this paper we presented \owl, a tool to automatically create 
Parsons problems in three tiers of increasing difficulty. \owl is 
built to deal with the challenging heterogeneity in computer science classrooms.
While internal differentiation addresses it effectively, it is time-consuming,
requiring additional lesson planning and material creation.
\owl can help to ease these efforts for educators by automatically
creating multiple Parsons problems differentiated by complexity from a sample
task or other parameters like programming concepts and targeted programming
language, utilising the capabilities of LLMs. Furthermore, student solutions 
for the created Parsons problems can be automatically checked by 
\owl, enabling immediate feedback.

Our study using
297 tiered sequences showed that $98.7\%$
of sequences achieved a positive complexity increase. Expert ratings of 60 generated
problem statements indicate high clarity and appropriateness for programming
beginners. A survey of 28 teachers praised the tool's time-saving potential and pedagogical value. Similarly, a study with 11 first-semester CS 
students reported a positive
experience highlighting supportive tiering and positive effects on 
self-efficacy.

Notably, as such a generator becomes better, repeated use becomes less necessary. \owl therefore 
stores each sequence with metadata, forming a reusable, queryable task library.

Future work includes: (i) adding keyboard and click-based input,
(ii) generating full worksheets from richer input,
(iii) integrating learning analytics,
(iv) offering finer teacher controls (e.g., construct limits, code length,
post-editing), and (v) enabling shared, curated task libraries.

To help the improvement of LLM-generated tiered Parsons problems and fully use 
its positive capabilities, a hosted version of \owl can be found at
\url{https://scratch.fim.uni-passau.de/codeowl/} and the project itself is 
publicly available at \url{https://github.com/se2p/codeowl}.